\documentclass[aps,pra,onecolumn,preprint,preprintnumbers,amsmath,amssymb,usenatbib,
nofootinbib,showkeys,showpacs,11pt]{revtex4}
\usepackage{graphicx}
\usepackage{dcolumn}
\usepackage{bm}
\usepackage{latexsym}
\usepackage{mathrsfs}
\usepackage{subfigure}
\usepackage[colorlinks]{hyperref}

\begin{document}

\title{Stability of viscous fluid in Bianchi type-VI model
with cosmological constant}
\author{J. Sadeghi$^{1}$}
\email{pouriya@ipm.ir}

\author{Ali R. Amani$^{2}$}
\email{a.r.amani@iauamol.ac.ir }

\author{N. Tahmasbi$^{1}$}
\email{N.Tahmasbi@stu.umz.ac.ir}
\affiliation{\centerline{$^1$Faculty of Basic Sciences, Departments of Physics, Mazandaran
University, }\\ P.O. Box 47415-416, Babolsar, Iran}

\affiliation{\centerline{$^2$Faculty of Basic Sciences, Department of Physics, Ayatollah Amoli Branch, Islamic Azad University,}\\ P.O. Box 678, Amol, Iran.}
\date{\today}



\begin{abstract}
In this paper, we investigate Bianchi type-VI cosmological model for the universe filled with dark energy and
viscous fluid in the presence of cosmological constant. Also,
we show accelerating expansion of the universe by draw-ing volume scale, pressure and energy density versus cosmic
time. In order to solve the Einstein's field equations, we assume the expansion scalar is proportional to a component of
the shear tensor. Therefore, we obtain the directional scale
factors and show the EOS parameter crosses over phantom
divided-line.
\end{abstract}
\maketitle


\section{Introduction}\label{s1}

Recent cosmological observations strongly indicate that our
universe has a accelerating expansion \cite{R1,R2}. The expansion of the universe
means increasing the metric distance between objects with
respect to time. The accelerating expansion with negative
pressure, called dark energy (DE). As we know, understand-ing of dark energy is greatest challenge of modern theoretical cosmology. In the standard model of cosmology, it is estimated that 73 percent of the total mass-energy of the universe occupy with dark energy. To explain mysterious dark
energy, a variety of the theoretical models have been pro-posed in the literature such as vacuum energy($\omega = -1$),
phantom($\omega < -1$), quintessence ($\omega > -1$), quintom (that
is the combination of phantom and quintessence), Chaplygin gas, tachyon and etc. Recently, there has been increasing interest for study Bianchi models \cite{R3, R4, R5, R6, R7, R8, R9, R10, R11}. On the other hand in cosmology, Bianchi
models describe a model of the universe which is homogeneous but not necessarily isotropic. we note here the observation of anisotropies in the Cosmic Microwave Background
(CMB) radiation and large scale wave-patterns are responsible for study anisotropic space-time. To consider more realistic models, we added the shear and bulk viscosity into
fluid. In cosmic physics, the cosmological constant play important role in expansion universe.\\
Bianchi type I universe with viscous fluid in the presence of $\Lambda$ term was investigated in \cite{R12}. The effect of bulk viscosity, with a time varying bulk
viscous coefficient and nonlinear spinor fields in BI universe filled with viscous fluid was studied by Refs. \cite{R13, R14}. The string-driven inflationary universe in
terms of effective bulk viscosity coefficient was achieved
by Ref. \cite{R15}. The density-dependent viscosity coefficient, Friedmann cosmology with a generalized equation of
state and bulk viscosity have shown in Refs. \cite{R16, R17}, also Bianchi type $VI_0$ cosmological models with viscous fluid presented in Ref. \cite{R18}. The magnetized Barotropic bulk viscous fluid
massive string in Bianchi type $VI_0$ have been studied by Refs. \cite{R19, R20}. So, all above information give us
motivation to investigate the dynamical effects of bulk and shear viscosity on the early evolution of the Bianchi type-VI model in the presence of a cosmological constant. The
paper is organized as follows: In section two and three we
introduce the metric background with basic equations and
obtain the solution of the field equations respectively. Also,
in section four and five we investigate some cosmological
parameters and study the effect of cosmological constant in
Bianchi type-VI Model. In section six, we discuss the stability conditions in early and late time, also we show some figures for the cosmological parameters. Finally, we have some
conclusion and further tasks for the corresponding model.


\section{ The metric background and basic equation}
\label{s2}
We start with homogeneous and anisotropic Bianchi type-VI line
element which is given by following background \cite{R21, R22, R23},
\begin{equation}\label{eq1}
  ds^2 = dt^2-a_1^2\,e^{-2mz}\,dx^2-a_2^2\,e^{2nz}\,dy^2-a_3^2\,dz^2,
\end{equation}
where the quantities $a_1, a_2$ and $ a_3$ are scale factors and $m,
n$ are some arbitrary constant. The Einstein's field equation is given
by
\begin{equation}\label{eq2}
 G_i ^j = R_i^j - \frac{1}{2} R\delta_i^j= k T_i^j,
\end{equation}
where $R$ is the Ricci scalar, $ R_i^j$  is the Ricci tensor, $k$ is
Enistein gravity constant (in gravitational units $\ {k=8\pi
G}/{c^4} = 1$ ), $ G_i ^j$ is the Enistein tensor, and $ T_i^j$  is
the energy-momentum tensor. This quantity in a four-dimensional
space has $ 4^2= 16 $ components . Here, the energy-momentum tensor
a viscous fluid will be as,
\begin{equation}\label{eq3}
T_i^j=(\rho + p_{eff})u_i u^j - p_{eff}  \delta_i^j + \eta
g^{j\beta} [u_{i; \beta} + u_{\beta; i} - u_i u^\alpha u_{\beta
;\alpha}- u_\beta u^\alpha u_{i ;\alpha}],
\end{equation}
where
\begin{equation}\label{eq4}
  p_{eff} = p + p_\xi + p_\eta  \,,   ( p_\xi = -3\xi H,  \,     p_\eta = 2\eta
  H).
\end{equation}
Here $\rho$ is the energy density of fluid, $u_\mu$ is the
4-velocity, $p$ is thermodynamical pressure, $ p_\xi$ and
$p_\eta$ the bulk and shear viscosity pressure are respectively,
$\eta$ is the coefficients of shear  viscosity, $\xi$ is the
coefficients of  bulk viscosity and the semicolon stands for
covariant differentiation. $\eta$ and $\xi$ are both positively
definite, i.e., $\eta > 0 $    $\xi > 0$. They may be either
constant or a function of time or energy, of course here we consider
the case with  $\eta ,\xi$ = constant. The equation of state
parameter $(\omega)$ is given by,
\begin{equation}\label{eq5}
 \omega = \frac {p}{\rho}
\end{equation}
According to the equation (\ref{eq3}) and by use of the metric
signature (+,-,-,-), so that $u^i$ = (+1,0,0,0) and $ u_i u^j$ = 1,
 we have
\begin{equation}\label{eq6}
T_0^0 = \rho
\end{equation}
\begin{equation}\label{eq7}
 T_1^1 = -p_{eff} + 2 \eta \frac {a\dot{}_1}{a_1}
\end{equation}
\begin{equation}\label{eq8}
T_2^2 = -p_{eff} + 2 \eta \frac {a\dot{}_2}{a_2}
\end{equation}
\begin{equation}\label{eq9}
T_3^3 = -p_{eff} + 2 \eta \frac {a\dot{}_3}{a_3}.
\end{equation}
 Hence the energy-momentum tensor has only diagonal elements and the rest of the elements are zero,
 \begin{equation}\label{eq10}
 T_i^j =  \left(
 \begin{array}{cccc}
 T_0^0 & 0 & 0 & 0 \\
 0 & T_1^1 & 0 &0 \\
 0 & 0 & T_2^2 & 0 \\
 0 & 0 & 0 & T_3^3 \\
 \end{array}
 \right)
 \end{equation}
The solution of Einstein's field equation (\ref{eq2}) for the line
element (\ref{eq1}) and equation (\ref{eq10}) lead us to have
following equations,
 \begin{equation}\label{eq11}
\frac {\dot{a_1} \dot{a_2} } {{a_1} {a_2}} + \frac{\dot{a_2}
\dot{a_3} } {{a_2} {a_3}} +\frac {\dot{a_1} \dot{a_3} } {{a_1}
{a_3}} - \frac {m^2-mn+n^2}{a_3^2} = k T_0^0
\end{equation}
\begin{equation}\label{eq12}
\frac{\ddot{a_2}}{a_2} + \frac{\ddot{a_3}}{a_3} + \frac {\dot{a_2}
\dot{a_3} } {{a_2} {a_3}} - \frac {n^2}{a_3^2} = k T_1^1
\end{equation}
\begin{equation}\label{eq13}
\frac{\ddot{a_3}}{a_3} + \frac{\ddot{a_1}}{a_1} + \frac {\dot{a_1}
\dot{a_3} } {{a_1} {a_3}} - \frac {m^2}{a_3^2} = k T_2^2
\end{equation}
\begin{equation}\label{eq14}
\frac{\ddot{a_1}}{a_1} + \frac{\ddot{a_2}}{a_2} + \frac {\dot{a_1}
\dot{a_2} } {{a_1} {a_2}} + \frac {mn}{a_3^2} = k T_3^3
\end{equation}
\begin{equation}\label{eq15}
m \frac {\dot{a_1}} {{a_1}} - n \frac {\dot{a_2}} {{a_2}} -
(m-n)\frac {\dot{a_3}} {{a_3}} = 0
 \end{equation}
where dot(.) denotes derivation with respect to cosmic time t. The
scale factor is a function of time which represents the relative
expansion of the universe. Here, average scale factor of Bianchi
type-VI metric is defined as,
\begin{equation}\label{eq16}
a = (a_1a_2a_3)^\frac{1}{3} .
\end{equation}
The volume of the universe in this model is given by
\begin{equation}\label{eq17}
V=a^3= a_1 a_2 a_3 .
\end{equation}
The expansion scalar can be written by,
\begin{equation}\label{eq18}
\theta = u_{;i}^i = \partial_i u^i +\Gamma_{i \alpha}^{i}u^\alpha = \frac{\dot{a_1}}{a_1}+ \frac{\dot{a_2}}{a_2}+
\frac{\dot{a_3}}{a_3}= \frac{\dot{V}}{V} ,
\end{equation}
where $\Gamma_{i \alpha}^{i}$ is the Christoffel symbol. And
shear scalar is given by,
\begin{equation}\label{eq19}
\sigma^2 = \frac{1}{2} \sigma_{ij} \sigma^{ij}
\end{equation}
where $\sigma_{ij}$ is,
\begin{equation}\label{eq20}
\sigma_{ij} = u_{i;j}- \frac{1}{2}(u_{i;\alpha} u^\alpha
u_j+u_{j;\alpha} u^\alpha u_i)- \frac{1}{3} \theta (g_{ij} - u_i
u_j) .
\end{equation}
And the diagonal elements of the shear tensor are given by,
\begin{equation}\label{eq21}
\sigma_1^1 = -\frac{1}{3}(-2\frac{\dot{a_1}}{a_1}+
\frac{\dot{a_2}}{a_2}+ \frac{\dot{a_3}}{a_3})=
\frac{\dot{a_1}}{a_1}- \frac{1}{3}\theta,
\end{equation}
\begin{equation}\label{eq22}
\sigma_2^2 =- \frac{1}{3}(-2\frac{\dot{a_2}}{a_2}+
\frac{\dot{a_1}}{a_1}+ \frac{\dot{a_3}}{a_3})=
\frac{\dot{a_2}}{a_2}- \frac{1}{3}\theta,
\end{equation}
\begin{equation}\label{eq23}
\sigma_3^3 =- \frac{1}{3}(-2\frac{\dot{a_3}}{a_3}+
\frac{\dot{a_2}}{a_2}+ \frac{\dot{a_1}}{a_1})=
\frac{\dot{a_3}}{a_3}- \frac{1}{3}\theta.
\end{equation}
The deceleration parameter in cosmology is a dimensionless quantity
that measure the cosmic acceleration of the expansion of universe.
This can be written as,
\begin{equation}\label{eq24}
q = -\frac{a \ddot{a}}{{\dot a}^2} =2-3\frac{V\ddot{V}}{\dot{V}^2} =
- (1+\frac{{\dot H}}{H^2} ) .
\end{equation}
The directional Hubble's parameters, and the mean Hubble parameter H
are respectively given as,
\begin{equation}\label{eq25}
H_1 = \frac{\dot{a_1}}{a_1} , \, H_2 = \frac{\dot{a_2}}{a_2} , \,
H_3 = \frac{\dot{a_3}}{a_3} , \,
\end{equation}
\begin{equation}\label{eq26}
H = \frac{\dot{a}}{a} = \frac{1}{3}\frac{\dot{V}}{V}.
\end{equation}
\section{Solution of the field equation}
Now we are going to obtain relation between scale factors which is obtained
by the equation (\ref{eq15}),
\begin{equation}\label{eq27}
(\frac{a_1}{a_3})^m = \kappa_1(\frac{a_2}{a_3})^n
\end{equation}
where $ \kappa_1 $ is constant of integration. In addition, we
assume that the expansion scalar in the model is proportional to a
component of the shear tensor, i.e.,
\begin{equation}\label{eq28}
\theta = N_3 \sigma_3^3,
\end{equation}
The structure of this equation and some condition in $\theta$ is
given by Ref. \cite{R24}. Inserting (\ref{eq28}) and
(\ref{eq18}) into (\ref{eq24}), the scale factor along $z$ direction
is,
\begin{equation}\label{eq29}
a_3 = N_0 V^{\frac{1}{3}+\frac{1}{N_3}}
\end{equation}
where, $ N_0 $ is also constant of integration. By using equation of
(\ref{eq17}), (\ref{eq27}) and (\ref{eq29}), the scale factors along
x,y directions are respectively,
\begin{equation}\label{eq30}
a_1 = \kappa_1^{\frac{1}{m+n}} N_0^{\frac{m-2n}{m+n}}
V^{\frac{1}{3}+\frac{m-2n}{N_3(m-n)}},
\end{equation}
\begin{equation}\label{eq31}
a_2 = \kappa_1^{-\frac{1}{m+n}} N_0^{\frac{n-2m}{m+n}}
V^{\frac{1}{3}+\frac{m-2n}{N_3(m-n)}}.
\end{equation}
so, we see that the scale factors express in terms of  $V$. By
subtracting equation of (\ref{eq13}) from (\ref{eq12}) and inserting
$a_1, a_2$ and $a_3$, finally we get following expression,
\begin{equation}\label{eq32}
\frac{\ddot{V}}{V} - \frac{N_3(m+n)^2}{3N_0^2
V^{\frac{2}{3}+\frac{2}{N_3}}} = k \frac {T_2^2 -
T_1^1}{3(m-n)/N_3(m+n)},
\end{equation}
where  $ \frac {T_2^2 - T_1^1}{3(m-n)/N_3(m+n)}= -2
\eta\frac{\dot{V}}{V}, $ so we have,
\begin{equation}\label{eq33}
\frac{\ddot{V}}{V} -
\frac{N_3(m+n)^2}{3N_0^2V^{\frac{2}{3}+\frac{2}{N_3}}} = -2k \eta
\frac{\dot{V}}{V}
\end{equation}
By multiplying both sides of above equation and with $ A_0 =
N_3(m+n)^2/3N_0^2 $, one can obtain,
\begin{equation}\label{eq34}
\ddot{V} - A_0 V^{ \frac{N_3-6}{3N_3}} = -2k \eta \dot{V}.
\end{equation}

\section{Some physical aspects of the models}
Now we check validity of this model with the drawing
of the physical quantities such as, p, $\rho$, $\omega$, $\theta$  and
$H$. By integrating from equation (\ref{eq34}), we obtain
\begin{equation}\label{eq35}
\dot{V} = -3k \eta V + \sqrt {9k^2 \eta^2 V^2 + A_1 V^{\frac
{4N_3-6}{3N_3}} + c_0}
\end{equation}
\begin{figure}[t]
\begin{center}
\includegraphics[scale=.3]{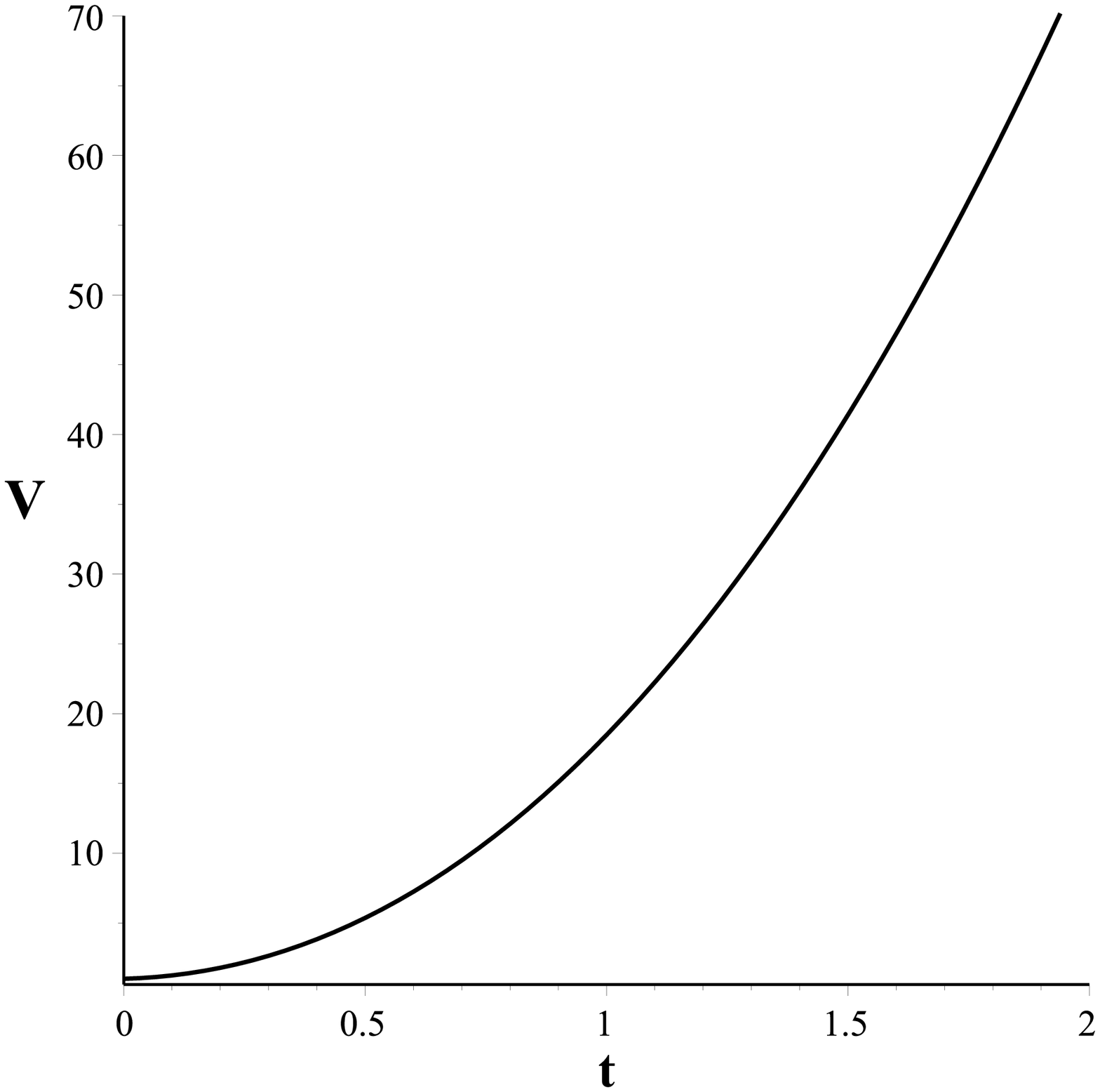}\includegraphics[scale=.3]{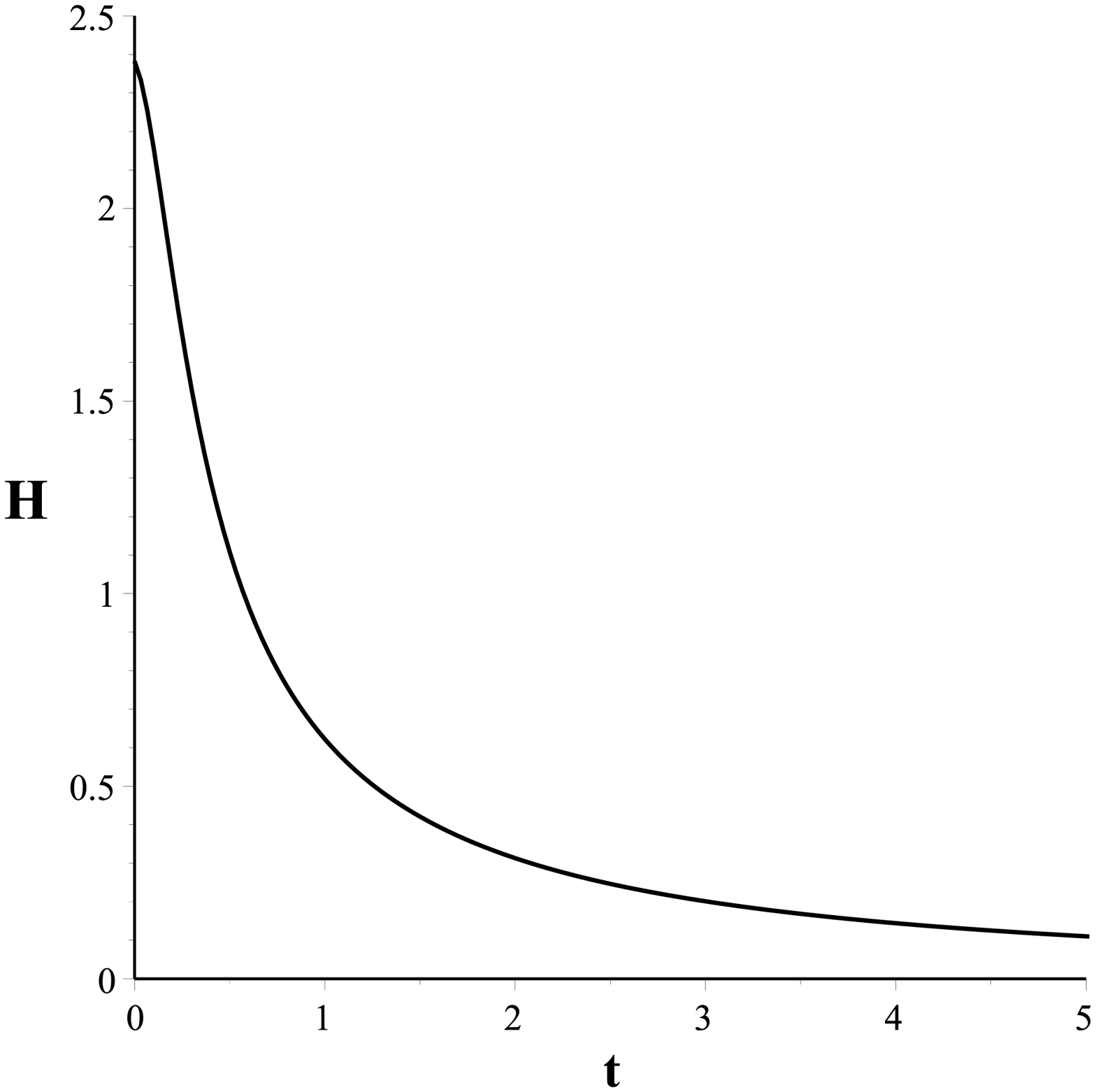}
\caption{The plots of the volume of the universe (left) and the
Hubble parameter (right) versus cosmic time for $ c_0=1,m=2,  n=4,
N_3=10 , N_0=2, k=\frac{2}{3},   \eta=0.1, \xi=0.05, V(0)=1,
\dot{V}(0)=1$  .}\label{fig1}
\end{center}
\end{figure}
where $c_0$ is  an integrating constant and $A_1 = \frac {3A_0
N_3}{2N_3-3}.$ We will find physical quantities in terms of V.
 By dividing (\ref{eq35}) into V the expansion scalar
will be following,
\begin{equation}\label{eq36}
\theta = \frac {\dot{V}}{V} = - 3 k \eta + \sqrt {9k^2 \eta^2 + A_1
V^{\frac {-(2N_3+6)}{3N_3}} + \frac{c_0}{V^2}}
\end{equation}
Here, also we can obtain from (\ref{eq26}) the mean Hubble parameter as,
\begin{equation}\label{eq37}
H = -k \eta + \frac {1}{3} \sqrt {9k^2 \eta^2 + A_1 V^{\frac
{-(2N_3+6)}{3N_3}} + \frac{c_0}{V^2}}
\end{equation}
Inserting $a_1$, $a_2$ and $a_3$ in Eq.(\ref{eq11}) and
Eqs. (\ref{eq5}-\ref{eq7}), we obtain the energy density as well as
the fluid pressure and EOS parameter respectively as follows,
\begin{equation}\label{eq38}
 \rho = \frac {1}{k} \left\{ \frac{X_1}{V^2} - X_2 V^{ \frac {-(2N_3+6)}{3N_3}} + X_3\left[ 18k^2\eta^2 -6k\eta \sqrt {9k^2 \eta^2 + A_1 V^{\frac {-(2N_3+6)}{3N_3}} + \frac{c_0}{V^2}}\right]\right\},
\end{equation}
\begin{figure}[t]
\begin{center}
\includegraphics[scale=.25]{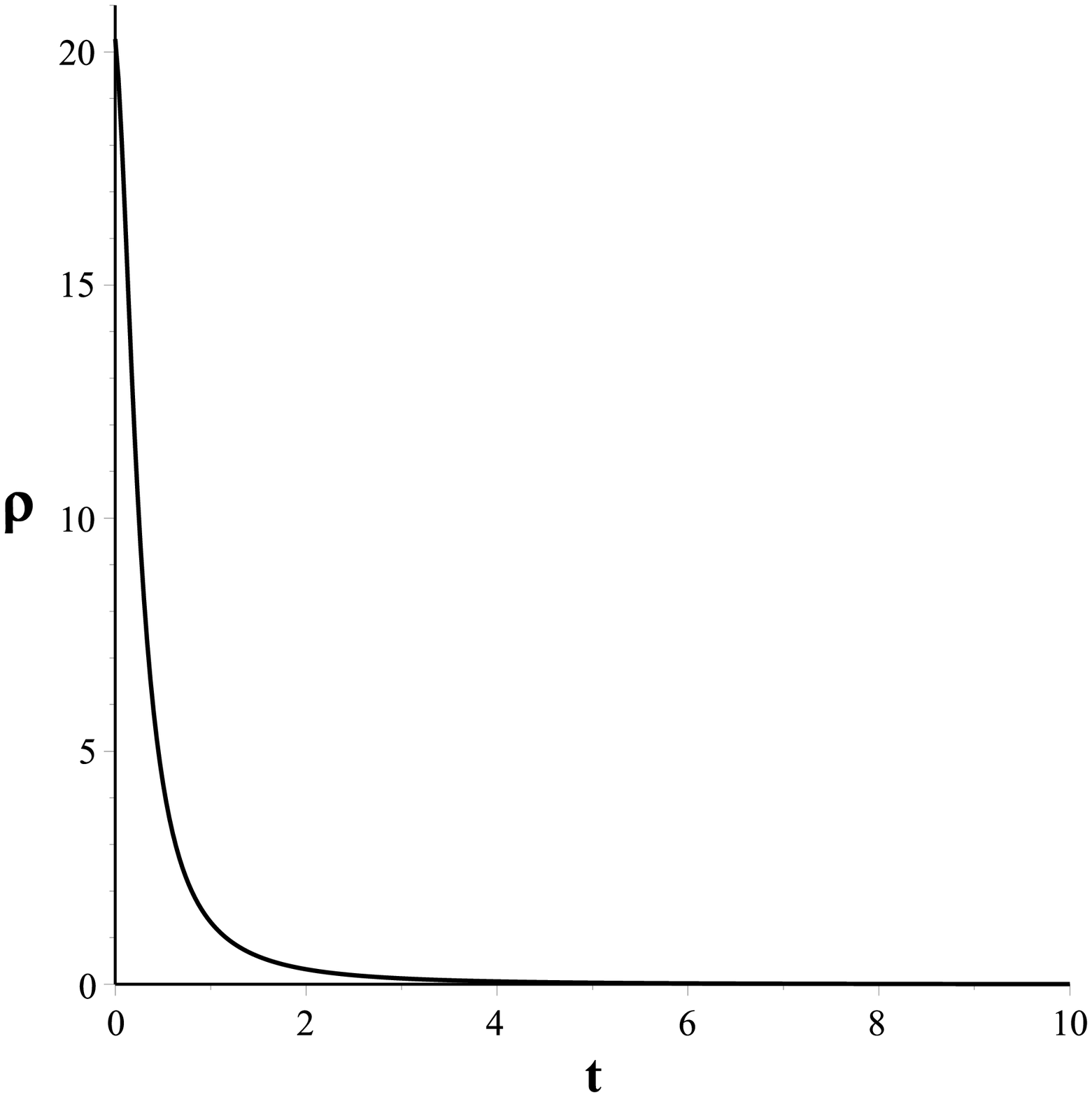}\includegraphics[scale=.25]{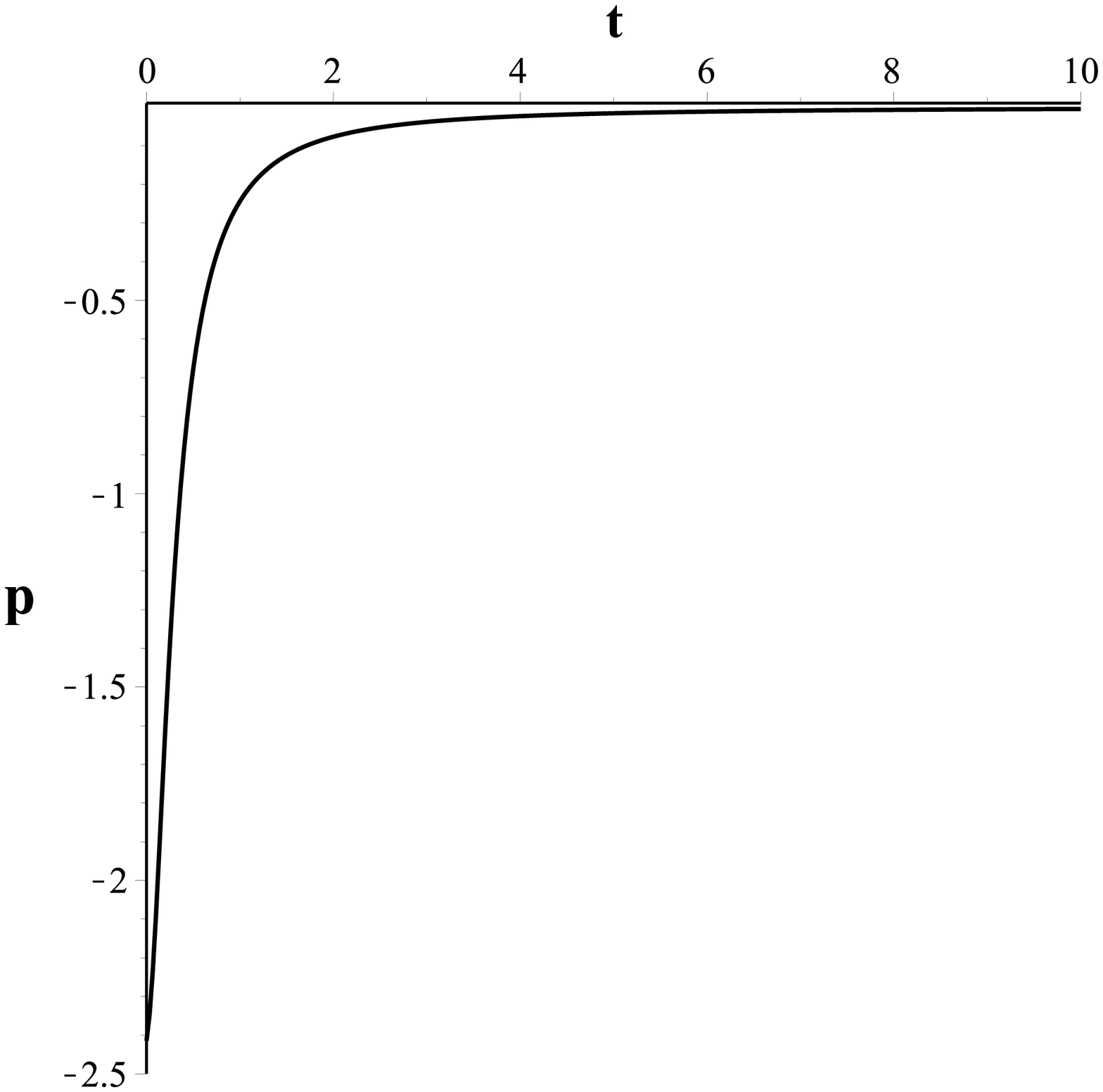}\includegraphics[scale=.25]{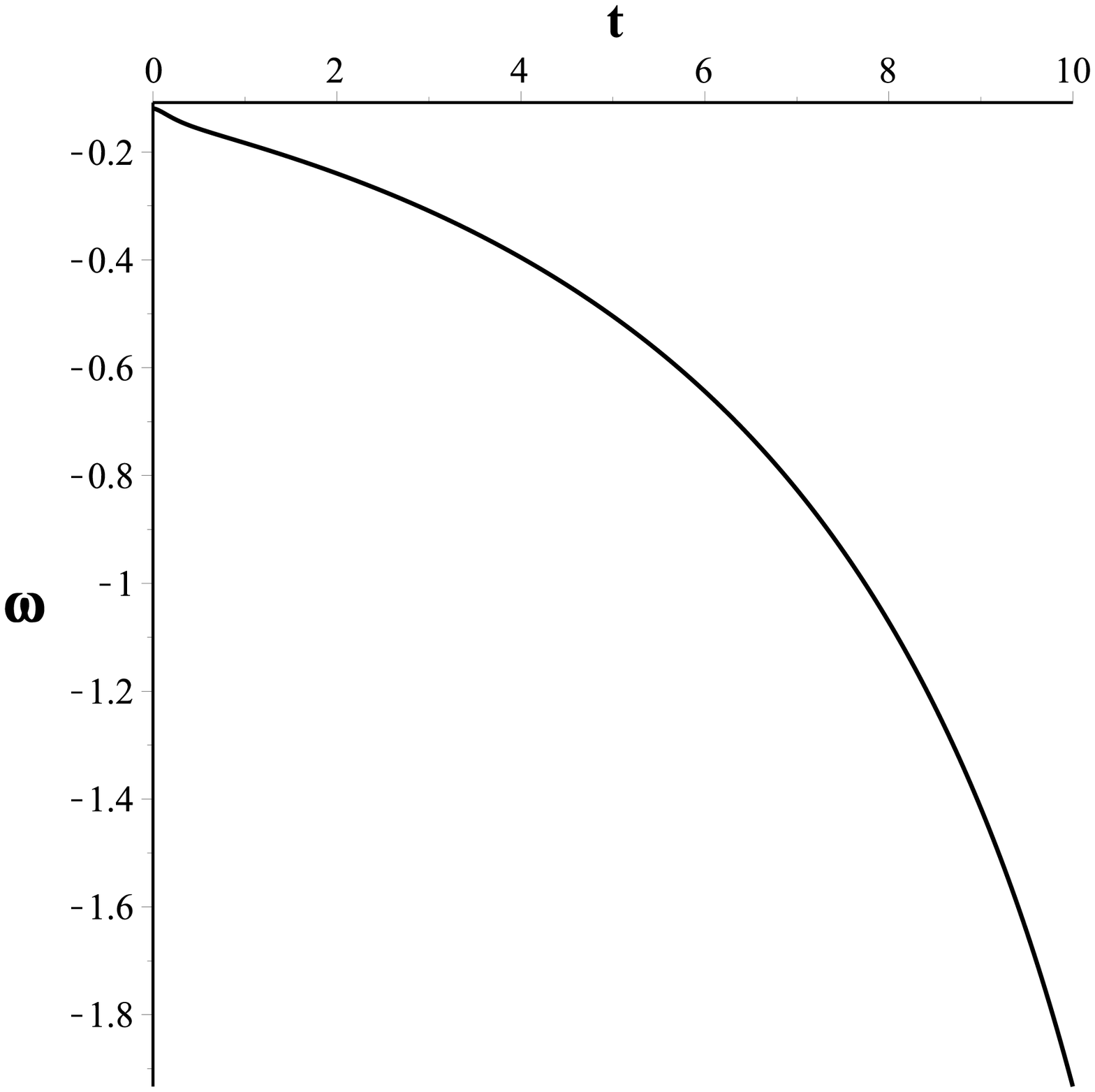}
\caption{The plots of the energy density (left), the pressure
(middle) and the EOS parameter(right) versus cosmic time for
$c_0=1,m=2,  n=4,  N_3=10 , N_0=2, k=\frac{2}{3},   \eta=0.1,
\xi=0.05, V(0)=1, \dot{V}(0)=1$  .}\label{fig2}
\end{center}
\end{figure}
\begin{eqnarray}\label{eq39}
\begin{aligned}
&p = \frac {1}{k} \left\{ \frac{X_1}{V^2} - X_4 V^{ \frac
{-(2N_3+6)}{3N_3}} + \left(3X_3-X_5\right)\left[ 6k^2\eta^2 -2k\eta
\sqrt {9k^2 \eta^2 + A_1 V^{\frac {-(2N_3+6)}{3N_3}} +
\frac{c_0}{V^2}}\right]\right\} \\&+
\left(\xi+X_6\eta\right)\left[-3k\eta +\sqrt {9k^2 \eta^2 + A_1
V^{\frac {-(2N_3+6)}{3N_3}} +\frac{c_0}{V^2}}\right].
\end{aligned}
\end{eqnarray}
\begin{equation}\label{eq40}
\omega = \frac{\begin{aligned}
&\left\{ \frac{X_1}{V^2} - X_4 V^{ \frac
{-(2N_3+6)}{3N_3}} + \left(3X_3-X_5\right)\left[ 6k^2\eta^2 -2k\eta
\sqrt {9k^2 \eta^2 + A_1 V^{\frac {-(2N_3+6)}{3N_3}} +
\frac{c_0}{V^2}}\right]\right\} \\&+
k\left(\xi+X_6\eta\right)\left[-3k\eta +\sqrt {9k^2 \eta^2 + A_1
V^{\frac {-(2N_3+6)}{3N_3}} +\frac{c_0}{V^2}}\right]
\end{aligned}}{ \left\{ \frac{X_1}{V^2} - X_2 V^{ \frac {-(2N_3+6)}{3N_3}} + X_3\left[ 18k^2\eta^2 -6k\eta \sqrt {9k^2 \eta^2 + A_1 V^{\frac {-(2N_3+6)}{3N_3}} + \frac{c_0}{V^2}}\right]\right\}}
\end{equation}
where, $$ X_1 = \left[\frac {1}{3} -3 \frac {m^2-mn+n^2}{N_3^2
(m+n)^2}\right]c_0 ,\qquad X_2 = \frac {m^2-mn+n^2}{N_0^2} - \frac
{X_1 A_1}{c_0} , \qquad X_3 = \frac{X_1}{c_0}$$ ,$$X_4=
\frac{2N_3-3}{3N_3} + \frac{mn}{N_0^2} - \frac{X_1 A_1}{c_0},\qquad
X_5 = \left[\frac {2}{3} + \frac {2n-m}{N_3(m+n)}\right],\qquad X_6=
2\frac{m-2n}{N_3(m+n)}.$$

We see in Fig. (\ref{fig2}), $\rho$  decreases with respect to time
and $p$  increases with respect to time. This  lead us to have
negative pressure and it is very large in early time. Also the EOS
parameter crosses over phantom divided-line ($\omega=-1$)( also
$\omega$ decreases with respect to time).

\section{ The cosmological constant}
\begin{figure}[h]
\begin{center}
\includegraphics[scale=.3]{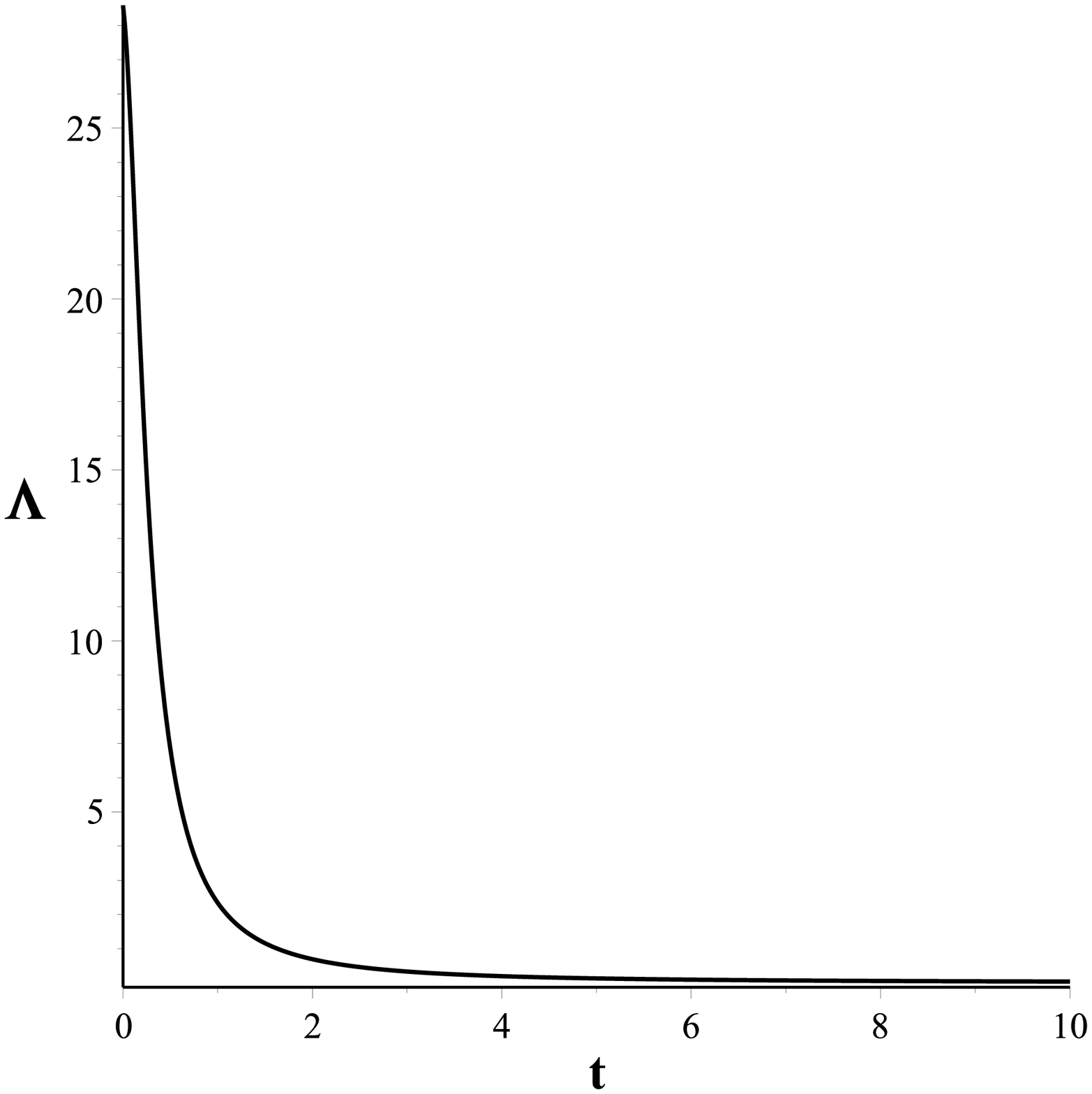}\includegraphics[scale=.3]{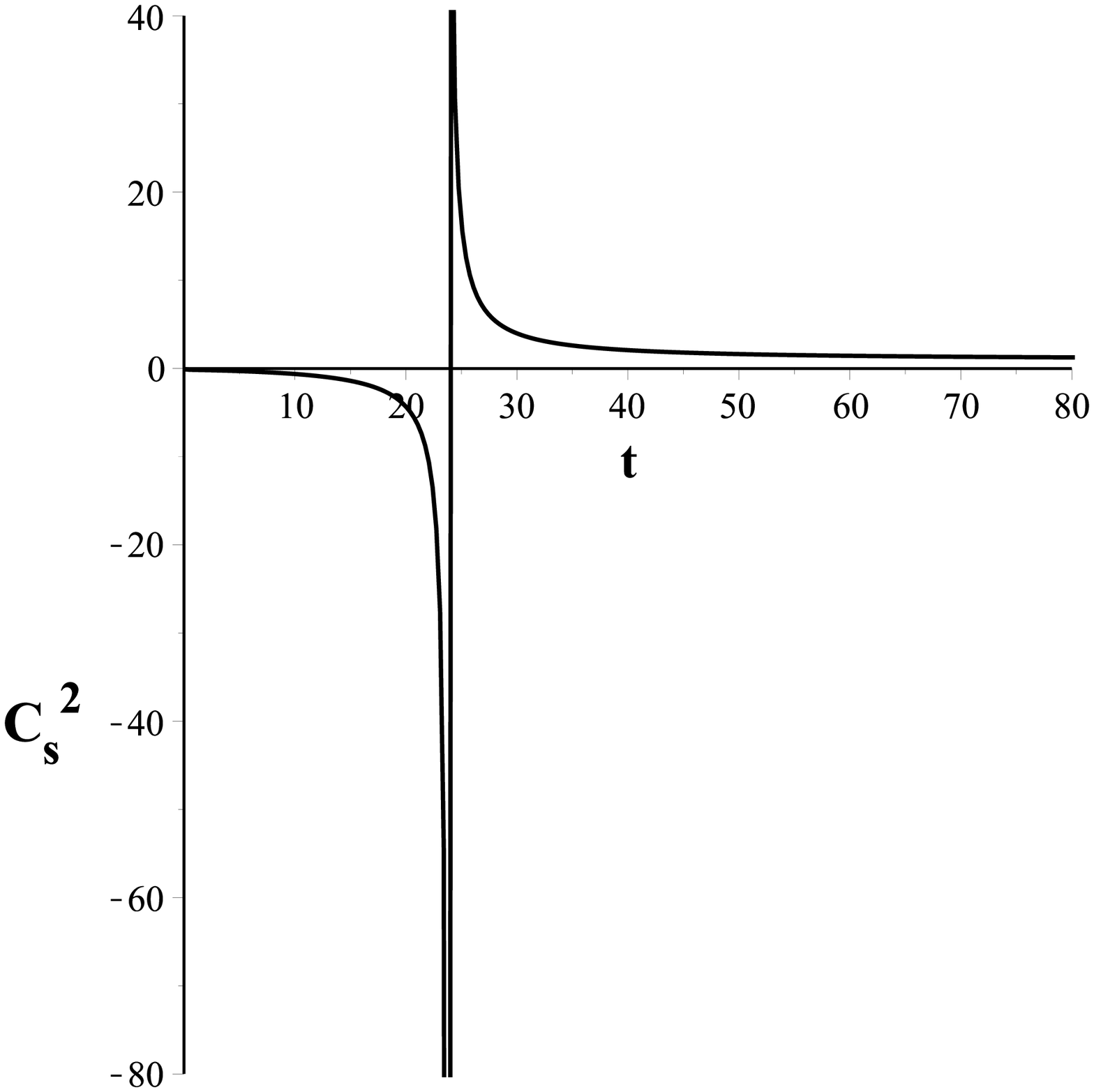}
\caption{The plots of the cosmological constant (left) and the
square of sound speed (right) versus cosmic time for $ c_0=1, m=2,
n=4, N_3=10 , N_0=2, k=\frac{2}{3}, \eta=0.1, \xi=0.05, V(0)=1,
\dot{V}(0)=1$.}\label{fig3}
\end{center}
\end{figure}

The simplest way to  introduce the dark energy is to add a
cosmological constant $\Lambda$ to the Riemann curvature scalar $R$
in the Lagrangian\cite{R25}. As we know,  the cosmological constant
can be interpreted as a measure of the vacuum energy density
\cite{R26}. Also, note here the cosmological constant depend on the
cosmic time directly or scale factor $a$ and  Hubble parameter $H$.
For example in Ref.\cite{R27} time-varying directly gravitational
and cosmological constant has been discussed. Recently, cosmological
observations by the Supernova cosmology project and the High-Z
Supernova search team suggest the existence of a positive
cosmological constant $\Lambda$ with the magnitude $\Lambda
(G\hbar/c^3)\approx 10^{-123}$. A positive, cosmological constant
accelerates the universal expansion with introducing a repulsive
force which can counterbalance the attractive force of gravity
leading to the static Einstein universe. On the other hand, we have
negative cosmological constant where ordinary matter tend to
decelerate it. Also the discussion of cosmological constant with
different of curvature  is given by Ref.\cite{R28}. So, we are going
to consider cosmological constant and  one can rewrite Eq.
(\ref{eq2}) as,
\begin{equation}\label{eq41}
R_i^j - \frac{1}{2} R\delta_i^j= k T_i^j - \Lambda \delta_i^j
\end{equation}
Here, we use the Einstein's equations and arrange the following
equations,
\begin{equation}\label{eq42}
\frac {\dot{a_1} \dot{a_2} } {{a_1} {a_2}} + \frac{\dot{a_2}
\dot{a_3} } {{a_2} {a_3}} +\frac {\dot{a_1} \dot{a_3} } {{a_1}
{a_3}} - \frac {m^2-mn+n^2}{a_3^2} = k T_0^0 - \Lambda
\end{equation}
\begin{equation}\label{eq43}
\frac{\ddot{a_2}}{a_2} + \frac{\ddot{a_3}}{a_3} + \frac {\dot{a_2}
\dot{a_3} } {{a_2} {a_3}} - \frac {n^2}{a_3^2} = k T_1^1 - \Lambda
\end{equation}
\begin{equation}\label{eq44}
\frac{\ddot{a_3}}{a_3} + \frac{\ddot{a_1}}{a_1} + \frac {\dot{a_1}
\dot{a_3} } {{a_1} {a_3}} - \frac {m^2}{a_3^2} = k T_2^2- \Lambda
\end{equation}
\begin{equation}\label{eq45}
\frac{\ddot{a_1}}{a_1} + \frac{\ddot{a_2}}{a_2} + \frac {\dot{a_1}
\dot{a_2} } {{a_1} {a_2}} + \frac {mn}{a_3^2} = k T_3^3- \Lambda
\end{equation}
The presence of cosmological constant lead us
to achieve the energy density and pressure as,
\begin{equation}\label{eq46}
\rho = \frac {1}{k} \left\{ \frac{X_1}{V^2} - X_2 V^{ \frac
{-(2N_3+6)}{3N_3}} + X_3\left[ 18k^2\eta^2 -6k\eta \sqrt {9k^2
\eta^2 + A_1 V^{\frac {-(2N_3+6)}{3N_3}} +
\frac{c_0}{V^2}}\right]+\Lambda \right\}
\end{equation}
and
\begin{eqnarray}\label{eq47}
\begin{aligned}
&p = \frac {1}{k} \left\{ \frac{X_1}{V^2} - X_4 V^{ \frac
{-(2N_3+6)}{3N_3}} + \left(3X_3-X_5\right)\left[ 6k^2\eta^2 -2k\eta
\sqrt {9k^2 \eta^2 + A_1 V^{\frac {-(2N_3+6)}{3N_3}} +
\frac{c_0}{V^2}}\right]+ \Lambda \right\} \\&+
\left(\xi+X_6\eta\right)\left[-3k\eta +\sqrt {9k^2 \eta^2 + A_1
V^{\frac {-(2N_3+6)}{3N_3}} +\frac{c_0}{V^2}}\right]
\end{aligned}
\end{eqnarray}
It is obvious that $\Lambda$ = cte only the diagrams $\rho$ and $p$
shifted. Now, we will show that the $\Lambda$ term decays with
respect to  time. In the framework of the $f(R, T )$ gravity, we can
get the cosmological constant as a function of the equation of state
parameter $(\omega)$ , the energy density $(\rho)$ and the trace of
the energy-momentum tensor (T). Of course, in following we consider
the cosmological constant $\Lambda$ as a function of the trace of
the energy-momentum tensor. In other words, the cosmological
constant in the gravitational Lagrangian is a function of the trace
of the energy-momentum tensor, and consequently the model was
denoted "$\Lambda(T)$ gravity". The dependence of the cosmological
constant $\Lambda$ on the trace of the energy-momentum tensor (T)
has been studied by \cite{R29,R30}. Therefore,  we can write
\begin{equation}\label{eq48}
\Lambda = trace (T_i^j) = T_0^0 +T_1^1+T_2^2+T_3^3,
\end{equation}
By using equations (\ref{eq6}-\ref{eq9}) and (\ref{eq4}), we obtain the
expression for $\Lambda $ as,
\begin{equation}\label{eq49}
\Lambda = \rho - 3p + 9\xi H
\end{equation}
We see in Fig. (\ref{fig3}) the variation of cosmological constant
with respect to time,  where $\Lambda$ is a decreasing function of
time $t$ and it approaches a small positive value at late time.

\section{The stability conditions}
 In this section, we are going to investigate the stability of the model.
We can consider the stability by using the function $c_s^2 =
dp/d\rho$. The stability condition occurs when the function $c_s^2$
becomes bigger than zero. The Fig. (\ref{fig3}) shows the behavior
of the speed of sound throughout the evolution of the universe. As
we see, there is instability in early time and there is stability in
late and future time.

\section{Conclusion}
In this paper, we have discussed the role of bulk and shear
viscosity in Bianchi type-VI cosmological model by assuming that $
\theta \propto \sigma$ in the presence of cosmological constant.
 The field equations have been solved exactly for the corresponding model.
Analytical solutions for the Hubble's parameter, energy density,
pressure, scale factor, volume scale and EOS parameter have been
derived. Finally, we checked our calculation with several figures.
For example in Fig. (\ref{fig1}), the volume of the universe and
Hubble parameter draw with respect to time and shown that during of
evolution universe they are  increasing and decreasing respectively.
Also we saw in Fig. (\ref{fig2}), $\rho$  decreases with respect to
time and $p$  increases with respect to time. This  lead us to have
negative pressure and it is very large in early time. Also the EOS
parameter crosses over phantom divided-line ($\omega=-1$)( also
$\omega$ decreases with respect to time). The behavior of
cosmological constant  with respect to cosmic time is given by Fig.
(\ref{fig3}). In (\ref{fig3}) we also observed that the $\Lambda$
term is initially infinite and it is decreasing function of time
that approaches to zero at late time. Also, the stability of such
system can be shown by Fig. (\ref{fig3}) which is plot of the square
of sound speed versus cosmic time.



\end{document}